\documentclass[article,12pt]{article} 
\usepackage{amscd,amsmath,amssymb,amsfonts,latexsym,mathrsfs,amsthm}
\usepackage{graphicx} 
\usepackage{setspace}  
\usepackage{graphics,epsfig}
\usepackage{sectsty}
\usepackage{natbib}
\usepackage[english]{babel}
\usepackage{amsmath}
\usepackage{amsfonts}
\usepackage{amssymb,amsthm}
\usepackage{bm}
\usepackage{mathrsfs}
\usepackage{caption}

\usepackage{subfigure}
\usepackage[usenames]{color}
\usepackage{rotating}
\usepackage{tabularx}
 \usepackage{algorithm,algorithmic}

\usepackage{flushend}
\usepackage{verbatim}
\usepackage{tabularx}
\usepackage{multirow} 

\usepackage{colortbl}

\usepackage{xcolor}

\definecolor{MYCOLOR0}{rgb}{0.92,0.92,0.92}
\definecolor{MYCOLOR}{rgb}{1,1,0}
\definecolor{MYCOLOR2}{rgb}{0.5,1,0.5}

\definecolor{MYCOLOR3}{rgb}{0.88,1,1}

\newcommand{\thetavec}{\bm{\theta}}

\newcommand{\x}{{\bf x}}
\newcommand{\y}{{\bf y}}
\newcommand{\z}{{\bf z}}
\newcommand{\X}{{\bf X}}
\newcommand{\E}{\mathrm{E}}
\newcommand{\Var}{\mathrm{Var}}

\newcommand{\Real}{\mathbb{R}}
\newcommand{\ESS}{\widehat{\text{ESS}}}
\newcommand{\MSE}{\text{MSE}}

\newcommand{\Rb}{\texttt{R2}}
\newcommand{\Rc}{\texttt{R3}}
\newcommand{\Na}{\texttt{N1}}

\newcommand{\Nc}{\texttt{N3}} 
\title{Advances in Importance Sampling}

\author{V\'ictor Elvira$^\star$ and Luca Martino$^\dagger$ \\
$^\star$  School of Mathematics, University of Edinburgh (United Kingdom)\\
$^\dagger$  Universidad Rey Juan Carlos de Madrid (Spain)
}

\date{}
 
\begin{document}

\setcitestyle{numbers}

\maketitle

\thispagestyle{empty}

\begin{abstract}
Importance sampling (IS) is a Monte Carlo technique for the approximation {of intractable} distributions and integrals with respect to them. The origin of IS dates from the early 1950s. {In the last decades, the rise of the Bayesian paradigm and the increase of the available computational resources} have propelled the interest in this theoretically sound methodology. In this paper, we first describe the basic IS algorithm and then revisit the recent advances in this methodology. We pay particular attention to two {sophisticated} lines. First, we focus on multiple IS (MIS), the case where more than one proposal is available. Second, we describe adaptive IS (AIS), the generic methodology for adapting one or more proposals.

\end{abstract} 

\section{Problem Statement}
 %%%%%%%%%%%%%%%%%%%%%%% %%%%%%%%%%%%%%%%%%%%%%% %%%%%%%%%%%%%%%%%%%%%
 In many problems of science and engineering intractable integrals must be approximated. Let us denote an integral of interest
 \begin{equation}
I(f)=E_{\widetilde \pi}[f(\x)]= \int f(\x)\widetilde{\pi}(\x) d\x,
\label{eq_integral}
\end{equation}
where $f:\Real^{d_x}\rightarrow \Real$, and $\widetilde \pi(\x)$ is a distribution of the r.v. $\X \in \Real^{d_x}$.\footnote{For the sake of easing the notation, from now on we use the same notation for denoting a random variable or one realization of a random variable.} Note that although Eq.~\eqref{eq_integral} involves a distribution, more generic integrals could be targeted with the techniques described below.

The integrals of this form appear often in the Bayesian framework, where a set of observations are available in $\y \in \mathbb{R}^{d_y}$, and the goal is in inferring some hidden parameters and/or latent variables $\x \in \mathbb{R}^{d_y}$ that are connected to the observations through a probabilistic model \cite{WileyLee14}. The information provided by the observations is compacted in the likelihood function $\ell(\y|\x)$ and the prior knowledge on $\x$ is encoded in the prior distribution $p_0(\x)$. Both sources of information are fused to create {through} the simple Bayes' rule the posterior probability density function (pdf), also called target distribution, given by
\begin{equation}
	\widetilde{\pi}(\x)= p(\x| \y)= \frac{\ell(\y|\x) p_0(\x)}{Z(\y)},
\label{eq:posterior}
\end{equation}
where $Z(\y)=\int \ell(\y|\x) p_0(\x) d\x $ is the marginal likelihood (a.k.a., partition function, Bayesian evidence, model evidence, or normalizing constant) \cite{Bernardo94,Robert07}.
In most models of {interest} $Z(\y)$ is unknown and in many applications it must be approximated \cite{Bernardo94,Box73,Robert07}. But even when its approximation is not needed, the unavailability of $Z(\y)$ implies that the posterior can be evaluated only up to that (unknown) constant, i.e., we can only evaluate 
\begin{equation}
\pi(\x)=\ell(\y|\x) p_0(\x),
\label{eq:target}
\end{equation}
that we denote as unnormalized target distribution.\footnote{From now on, we drop $\y$ to ease the notation, {e.g., $Z \equiv Z(\y)$.}} Table \ref{tab:notation} summarizes the notation of this article.

\begin{table}[!t]
\centering
\caption{Summary of the notation.}
\vspace{0.1cm}
	\begin{tabular}{|c|l||c|l|}
    \hline
 \cellcolor{MYCOLOR0} $d_x$ & \multicolumn{3}{l|}{dimension of the inference problem, $\x\in\mathbb{R}^d_x$} \\ 
 \cellcolor{MYCOLOR0} $d_y$ & \multicolumn{3}{l|}{dimension of the observed data, $\y\in\mathbb{R}^d_y$} \\ 
\hline
\hline
 \cellcolor{MYCOLOR0}$\x$ &\multicolumn{3}{l|}{r.v. of interest; parameter to be inferred}\\
 \cellcolor{MYCOLOR0}$\y$ &\multicolumn{3}{l|}{observed data} \\
  \hline
  \hline
  \cellcolor{MYCOLOR0} $\ell(\y|\x)$ & \multicolumn{3}{l|}{likelihood function} \\ 
  \cellcolor{MYCOLOR0} $p_0(\x)$ & \multicolumn{3}{l|}{prior pdf} \\ 
  \cellcolor{MYCOLOR0} $\widetilde{\pi}(\x)$ & \multicolumn{3}{l|}{posterior pdf (target), $\widetilde{\pi}(\x)\equiv p(\x| \y)$} \\
  \cellcolor{MYCOLOR0} $\pi(\x)$ & \multicolumn{3}{l|}{posterior density function (unnormalized target) , $\pi(\x) \equiv \ell(\y|\x) g(\x)\propto {\bar \pi}(\x)$} \\
    \cellcolor{MYCOLOR0} $q(\x)$ & \multicolumn{3}{l|}{ {proposal density}} \\
      \cellcolor{MYCOLOR0} $Z$ & \multicolumn{3}{l|}{ {normalizing constant or marginal likelihood, $Z \equiv Z(\y)$}} \\
      %   \hline 
    \cellcolor{MYCOLOR0} $I(f)$ & \multicolumn{3}{l|}{integral to be approximated, $I(f)\equiv E_{\widetilde \pi}[f(\x)]$.} \\
   \hline
\end{tabular} 
\label{tab:notation}
\end{table}

The integral $I(f)$ cannot be computed in a closed form in many practical scenarios and hence must be approximated. The approximation methods can be {divided into} either deterministic or stochastic. While {many} deterministic numerical methods are available in the literature \cite{Acton90,WileyAusin14_2,Burden00,Kythe04,Plybon92}, it is in general accepted that they tend to become less efficient than stochastic approximations when the problem dimension $d_x$ grows.

%%%%%%%%%%%%%%%%%%%%%%%%%%%%%%%%%%%%%%%%
\subsection{Standard Monte Carlo integration}
%%%%%%%%%%%%%%%%%%%%%%%%%%%%%%%%%%%%%%%%

The Monte Carlo approach consists in approximating the integral $I(f)$ in Eq. \eqref{eq_integral} with random samples \citep{Dunn2011,Jaeckel02,Gentle04,Kroese11,Liu04b,WileyMedova08,Robert04,luengo2020survey}. In the standard Monte Carlo solution (often called instinctively vanilla/raw/classical/direct Monte Carlo), $N$ samples $\x_n$ are independently simulated from $\widetilde \pi(\x)$. The standard Monte Carlo estimator is built as
\begin{equation}
\label{IdealMC}
 {\overline I}^N(f)=\frac{1}{N} \sum_{t=1}^N f(\x_n).
%\xrightarrow[]{p}
\end{equation}
First, note that ${\overline I}^N(f)$ is unbiased since $E_{\widetilde \pi}[ {\overline I}^N(f)]=I(f)$. Moreover, due to the weak law of large {numbers}, it can be shown that ${\overline I}_N$ is consistent and then converges in probability to the true value $I$, i.e., ${\overline I}^N(f) \overset{p}{\longrightarrow}I(f)$, which is equivalent to {stating} that, for any positive number $\epsilon>0$, we have $\lim\nolimits_{N\rightarrow \infty}\mbox{Pr}( |{\overline I}^N(f)-I(f) |> \epsilon)=0$. 
The variance of $ {\overline I}^N(f)$ is simply 
$\overline \sigma^2 = \frac{1}{N}\left(I(f^2) - I(f)^2 \right)$.  
If the second moment is finite, $I(f^2)<\infty$, then the central limit theorem (CLT) applies and the estimator converges in distribution to a well-defined Gaussian when $N$ grows to infinity i.e., 
\begin{equation}
 \sqrt{N} \left( {\overline I}^N(f) - I(f) \right) \overset{d}{\longrightarrow} \mathcal{N}(0,\overline \sigma^2).
\end{equation}
There exist multiple families of Monte Carlo methods \cite{Robert04,robert2014m,taimre2019monte}. We address the interested reader to the articles in Markov chain Monte Carlo (including Metropolis-Hastings \cite{robert2015,martino2014metropolis} and Gibbs sampling \cite{christen2014g}) and previous articles in importance sampling \cite{wang2014importance,kong2014importance}.  

\section{Importance sampling}

\subsection{Origins}

The first use of the importance sampling (IS) methodology dates from 1950 for rare event estimation in statistical physics, in particular for the approximation of the probability of nuclear particles penetrating shields \cite{kahn1950random}. IS was later used as a variance reduction technique when standard Monte Carlo integration was not possible and/or not efficient {\cite{Hesterberg95}}. The {renewed} interest in IS has run in parallel with the hectic activity in the community of Bayesian analysis and its ever increasing computational demands. In most cases, the posterior in \eqref{eq:posterior} is not available due to the intractability of the normalizing constant. 
See \cite{tokdar2010importance} for a previous review in IS.

\subsection{Basics}

Let us start defining the proposal pdf, $q(\x)$, used to simulate the samples. It is widely accepted that the proposal is supposed to have heavier tails than the target, i.e., the target $\widetilde \pi(\x)$ decays faster than $q(\x)$ when $\x$ is \emph{far} from the region where most of the probability mass is concentrated. However, this usual restriction is too vague and it will be clarified below. Here, we simply stick to the restriction that $q(\x)>0$ for all $\x$ where $\widetilde \pi(\x)f(\x) \neq 0$. 
IS is constituted of two simple steps:

\begin{enumerate}
  \item \textbf{Sampling}: $N$ samples are simulated as
\begin{equation} 
\x_n \sim q(x),\qquad n=1,...,N.
\label{sampling_static}
\end{equation} 

\item \textbf{Weighting}: Each sample receives an associated importance weight given by
\begin{equation} 
  w_n= \frac{\pi(\x_n)}{{q(\x_n)}}, {\quad n=1,\ldots,N.}
\label{weighting_static}
\end{equation} 
\end{enumerate}
The importance weights {describe} how representative the samples simulated from $q(\x)$ are when one is interested in computing integrals w.r.t. $\widetilde \pi(\x)$. 
The set of $N$ weighted samples can be used to approximate the generic integral $I(f)$ of Eq. \eqref{eq_integral} by the two following IS estimators: 
\begin{itemize}
  \item Unnormalized (or nonnormalized) IS (UIS) estimator:
  \begin{equation}
  \widehat{I}^N(f) = \frac{1}{NZ}  \sum_{n=1}^N w_n f(\x_n).
\label{eq_UIS}
\end{equation}
Note that the UIS estimator can be used only when $Z$ is known.

  \item Self-normalized IS (SNIS) estimator:
\begin{equation}
  \widetilde{I}^N(f) =  \sum_{n=1}^N \overline w_n f(\x_n),
\label{eq_SNIS}
\end{equation}
where 
\begin{equation}
  \overline w_n = \frac{w_n}{\sum_{j=1}^N w_j}
\label{eq_SNIS}
\end{equation}
are the normalized weights.  
\end{itemize}

The derivation of the SNIS estimator departs from the UIS estimator of Eq. \eqref{eq_UIS}, substituting $Z$ by its unbiased estimator \cite{Robert04}
\begin{equation}
\widehat{Z}=\frac{1}{N}\sum_{n=1}^N w_n.
\label{eq_Z_estimator}
\end{equation}
 {After a few manipulations, one recovers} Eq. \eqref{eq_SNIS}. The normalized weights also allow to approximate the target distribution by
\begin{eqnarray}
\widetilde{\pi}^N(\x) &=& \sum_{n=1}^N \overline{w}_n\delta(\x-\x_n),
\label{approximation}
\end{eqnarray}
where $\delta$ represents the Dirac measure.

\subsection{Theoretical analysis}
The UIS estimator is unbiased since it can be easily proven that $\E_q[\widehat{I}^N(f)] = I(f)$. Its variance $\E_q[\widehat{I}^N(f)] = \frac{\sigma_q^2}{N}$ is given by
\begin{equation}
\sigma_q^2 = \int \frac{\left( f(\x)\widetilde \pi(\x) - I(f)q(\x) \right)^2}{q(\x)}d\x,
\label{eq_onesample_IS_variance}
\end{equation}
if $q(\x)>0$ for all $\x$ where $\widetilde \pi(\x)f(\x) \neq 0$, as we have stated above \cite{owen2013monte}. We remark that it is not strictly necessary to have a proposal with {heavier tails} than the target distribution as long as $\sigma_q^2<\infty$. One counter example is a case where $f(\x)$ decays fast enough to compensate the heavier tails of the target distribution. Another counter example is a case where $f(\x)$ takes non-zero and finite values only in a bounded set.

Note that $q(\x)$ is chosen by the practitioner and a good choice is critical for the efficiency of IS. Let us first suppose that $\text{sign}\left(f(\x)\right)$ is constant for all $\x$ and $I(f)\neq0$. Let us also suppose that it is possible to simulate from
 \begin{equation}
q^*(\x) = \frac{f(\x)\widetilde \pi(\x)}{\int f(\z)\widetilde \pi(\z)d\z}.
\label{eq_opt_proposal}
\end{equation}
Then, the UIS estimator, {for any $N\geq 1$ number of samples}, yields a zero-variance unbiased estimator, since the numerator in \eqref{eq_onesample_IS_variance} is zero, 
 and hence $\sigma_q^2=0$. However, it is very unlikely to have access to the proposal of \eqref{eq_opt_proposal}. The main reason is that its normalizing constant is exactly the intractable integral we are trying to approximate, $I(f)$. However, $q^*(\x)$ gives the useful intuition that the proposal should have mass proportional to the targeted integrand in Eq. \eqref{eq_integral}. More precisely, inspecting \eqref{eq_onesample_IS_variance}, we see that the efficiency is penalized with the mismatch of $f(\x)\widetilde \pi(\x)$ and $q(\x)$, with this penalization amplified inversely proportional to the density $q(\x)$. This explains the usual safe practice of over-spreading the proposal. The case where $\text{sign}\left(f(x)\right)$ alternates can be easily modified by splitting the function as $f(x) = f_+(x)+f_-(x)$, where $f_+(x)$ is non-negative and $f_-(x)$ is non-positive. It is easy to show that with the use of two proposals and $N=2$, a zero-variance estimator is possible (see \cite[Section 9.13]{owen2013monte}).
In summary, the UIS estimator, $ \widehat{I}^N(f)$ is unbiased, while the  $ \widetilde{I}^N(f)$ is only asymptotically unbiased, i.e., with a bias that goes to $0$ when $N$ grows to infinity.
Both UIS and SNIS are consistent estimators of $I$ with a variance that depends on the discrepancy between $\pi(\x)|f(\x)|$ and $q(\x)$, although the variance of the SNIS is more difficult to evaluate and its bias place also a central role when $N$ is not large enough \cite{owen2013monte}. 

%{\color{red}
When several different moments $f$ of the target must {be} estimated, a common strategy in IS is to decrease the mismatch between the proposal $q(\x)$ and the target $\widetilde \pi(\x)$ \cite{doucet2009tutorial}. This is equivalent to minimizing the variance of the weights and consequently the variance of the estimator $\widehat{Z}$, and it is closely linked to the diagnostics of Section \ref{sec_diagnostics}.

\subsection{Diagnostics}
\label{sec_diagnostics}
 It is a legitimate question to wonder {about} the efficiency of the set of simulated weighted samples in the task of approximating the target distribution and/or moments of it. Usual metrics of efficiency involve the computation of the variance of the IS estimators. However, the computation of those variances is intractable, and even more, their approximation is usually a harder problem than computing Eq. \eqref{eq_integral} (see \cite[Chapter 9.3]{owen2013monte} for a discussion). A classic diagnostic metric in the IS literature \cite{Kong92} is  
 \begin{equation}
\ESS = \frac{1}{\sum_{n=1}^N \bar w_n^2}.
\label{eq_rule_of_thumb}
\end{equation}
Note that $1 \leq \ESS\leq N$, taking the value $\ESS = 1$, when one $w_j=1$ and hence $w_i=0$, for all $i\neq j$. Therefore, $\ESS = N$ only when $w_j = 1/N$ for all $j=1,...,N$. Hence $\ESS$ measures the discrepancy among normalized weights. This diagnostic is commonly called \emph{effective sample size}, although it is an approximation of {the} more reasonable but intractable diagnostic given by \cite{elvira2018rethinking}
\begin{equation}
\text{ESS}^* = N\frac{\Var[\overline I^N]}{\MSE[\widetilde I^N]}. 
\label{eq_ess_mse}
\end{equation} 
Then, $\text{ESS}^*$ can be interpreted as the number of standard Monte Carlo that are necessary to obtain the same performance (in terms of MSE) as with the SNIS estimator with $N$ samples. The interested reader can find the derivation from $\text{ESS}^*$ to $\ESS$ through a series of approximations and assumptions that rarely hold (see \cite{elvira2018rethinking} for a thorough analysis). In practice, a low $\ESS$ is a symptom of malfunctioning, but a high $\ESS$ does not necessarily {imply good behavior} of the IS method. 

New ESS-like methods have been proposed in the last years. In \cite{martino2017effective,martino2016alternative}, novel discrepancy measures with similar properties to $\ESS$ are proposed and discussed, mitigating some of the deficiencies of the original diagnostic. For instance, an alternative to $\text{ESS}^*$ is {using $1/\max(\bar w_n)$ instead}, which preserves some of those properties (e.g., it takes values between $1$ and $N$, being $1$ if all the normalized weights are zero except one, and $N$ if all weights are the same). Another metric in the same spirit has been recently proposed in \cite{huggins2019sequential}. {Finally, the use of the importance trick within quadrature schemes has been recently proposed \cite{elvira2019gauss,elvira2020importance}. Note that these \emph{importance quadrature} schemes are not stochastic but strongly inspired in IS and its variants.} 

\subsection{Other IS {schemes}}

The research in IS methods has been very active in the last decade not only in the development of novel methodology but also for increasing the understanding and the theoretical behavior of IS-based methods. For instance, \cite{agapiou2017importance} unifies different perspectives about how many samples are necessary in IS for a given proposal and target densities, a problem that is usually related to some notion of distance (more precisely divergence) between the two densities. With a similar aim, in \cite{chatterjee2018sample} it is shown that in a fairly general setting, IS requires a number of samples proportional to the exponential of the KL divergence between the target and the proposal densities. The notion of divergences between both densities is  also explored in \cite{ryu2014adaptive} through the R\'enyi generalized divergence, and in \cite{miguez2017performance} in terms of the Pearson $\chi^2$ divergence. Both divergences are connected with the variance of the $\widehat{Z}$ estimator in Eq. \eqref{eq_Z_estimator}.

\subsubsection{Transformation of the importance weights}
\label{sec_crop}
 As described in Section \ref{sec_diagnostics}, a large variability in the importance weights is {usually responsible for a} large variance in the IS estimators. One alternative is adapting the proposals in order to diminish the mismatch with the target, as we describe in Section \ref{sec_ais}. However, this usually means throwing away past weighted samples (or stick to large variance estimators from the early iterations). Another alternative is the nonlinear transformation of the IS weights. The first work in this line is the \emph{truncated importance sampling}  \cite{ionides2008truncated} where the standard unnormalized weights $w_n$ are truncated as $w_n' = \min(w_n,\tau)$, where $\tau$ is a maximum value allowed for the transformed/truncated weights. The consistency of the method and a central limit theorem of the modified estimator are proved. This transformation of the weights was also proposed in \cite{koblents2015population}, and called \emph{nonlinear importance sampling} within an adaptive IS scheme (N-PMC algorithm). The convergence of this method is analyzed in  \cite{koblents2015population,miguez2017performance,miguez2018analysis}. The underlying problem that those methods fight is the right {heavy tail} in the distribution of the importance weights when the proposal is not well fit. In \cite{vehtari2015pareto}, the authors go a step beyond by characterizing the  distribution of the importance weights with generalized Pareto distribution that fits the upper tail. Based on this fitting, a method is proposed for the stabilization of the importance weights. The authors provide proofs for consistency, finite variance, and asymptotic normality. See \cite{martino2018comparison} for a review of the clipping methodologies.

\subsubsection{Particle filtering (sequential Monte Carlo)}
Particle filtering (also known as sequential Monte Carlo) is an {IS-based} methodology for performing approximate Bayesian inference on a hidden state that evolves over the time in state-space models, a class of probabilistic Markovian models. Due to the structure of the Bayesian network, it is possible to process sequentially and efficiently the observations related to the hidden state for building the sequence of filtering distributions (i.e., the posterior distribution of a given hidden state conditioned to all available observations). Particle filters (PFs) are based on importance sampling, incorporating in most cases a resampling step that helps to increase the diversity of the particle approximation \cite{Douc05,li2015resampling}. Since the publication of the seminal paper \cite{Gordon93} where the bootstrap PF is developed (BPF), a plethora of PFs have been proposed in the literature \cite{doucet2000rao,Pitt01,Kotecha03a,djuric2007multiple,elvira2018search}. Advanced MIS and AIS techniques are often implicit in those algorithms, but they are rarely explicit. In \cite{elvira2019elucidating}, a novel perspective of BPF and auxiliary PF (APF) based on MIS is introduced, and in \cite{branchini2020optimized}, the perspective is exploited for optimizing the mixture proposal. {In these state-space models, the ESS and its approximations are also used as diagnostics metrics for PF (see Section \ref{sec_diagnostics}). Moreover, since the observations are dependent in these models, other metrics have been recently developed, mostly based on the predictive distribution of the observations \cite{Lee15,Bhadra16,elvira2017adapting,elvira2019new}.}

\section{Multiple importance sampling (MIS)}

The IS methodology can be easily extended when the samples are simulated from $M$ proposals, $\{q_m(\x)\}_{m=1}^M$, instead of only one. In a generic setting, one can consider that $n_m$ samples are simulated from each proposal ($\sum_{j=1}^m n_j = 1$) and weighted appropriately. This extension is usually called multiple importance sampling (MIS), and it has strong connections with the case of standard IS with a single mixture proposal with components that are distributions, which is sometimes called mixture IS. Here we consider mixture IS as a subset of MIS methods when $n_m$ are not deterministic number of samples but r.v.'s instead.

\subsection{Generalized MIS}
\label{sec_gen_mis}
A unifying framework of MIS has been recently proposed in \cite{elvira2019generalized}. {The framework} encompasses most of existing IS methods with multiple proposals, proposes new schemes, and {compares} them in terms of variance. For the sake of clarity, the framework is described in the case where (a) no prior information about the adequateness of the proposals is available; and (b) $M=N$ proposals are available (i.e., exactly the same number of proposals than samples to be simulated). However, straightforward extensions are possible to more generic settings.
According to this framework, a MIS is proper if it fulfills two conditions related to the sampling and weighting processes. A valid sampling scheme for the simulation of $N$ samples, $\{\x_n \}_{n=1}^N$, can be agnostic to the dependence of those samples but must fulfill the following {statistical} property: a sample $\x$ randomly picked from the whole set of $N$ simulated samples must be distributed as the mixture of proposals {$\psi(\x) = \frac{1}{M}\sum_{m=1}^M q_m(\x)$}. A valid weighting scheme must yield an unbiased and consistent UIS estimator, $\widehat I^N$. 
{These properness conditions extend the standard properness in IS established by \cite{Liu04b}, and have been also used to assign proper importance weights to resampled particles \cite{martino2016weighting}.}
The paper analyzes and ranks several resulting MIS schemes (different combination of valid sampling and weighting procedures) in terms of variance. Due to space restrictions, here we show only two MIS schemes commonly used in the literature.
Let us simulate exactly one sample per proposal (sampling scheme $\mathcal{S}_3$ in \cite{elvira2019generalized}) as  
\begin{equation}
\x_n \sim q_n(\x),\qquad n=1,...,N.
\label{eq_sampling_mis}
\end{equation}
The next two weighting schemes are possible (among many others):
\begin{itemize}
\medskip
\item \textbf{Option 1}: Standard MIS (s-MIS, also called $\Na$ scheme): 
\begin{equation} w_n= \frac{\pi({\bf x}_n)}{{q_n({\bf x}_n)}}, \quad n=1,\ldots, N.  \label{weights_SIS}
\end{equation}
\medskip
\item \textbf{Option 2}: Deterministic mixture MIS (DM-MIS, also called $\Nc$ scheme):
\begin{equation} 
  w_n=\frac{\pi({\bf x}_n)}{\psi({\bf x}_n)}=\frac{\pi({\bf x}_n)}{\frac{1}{N}\sum_{j=1}^{N}q_j({\bf x}_n)}, \quad n=1,\ldots, N.\nonumber
\label{f_dm_weights_static}
\end{equation}
\end{itemize}
In both cases, it is possible to build the UIS and SNIS estimators. In \cite{elvira2019generalized}, it is shown that 
$${\Var[\widehat I_{\Nc}^N]\leq \Var[\widehat I_{\Na}^N]},$$
i.e., that using the second weighting option with the whole mixture in the denominator is always better than using just the proposal that simulated the sample (the equality in the variance relation happens only when all the proposals are the same). The result is relevant since ${\Na}$ is widely used in the literature but it should be avoided whenever possible. Note that both $\Na$ and $\Nc$ require just one target evaluation per sample. However, $\Nc$ requires $N$ proposal evaluations per sample, while $\Na$ just one. {For a small number of proposals}, or when the target evaluation is very expensive (and hence the bottleneck), this extra complexity in $\Nc$ may be not relevant, but it can become cumbersome otherwise. Several MIS strategies have been proposed in the literature to alleviate this problem. In \cite{elvira2015efficient}, a partition of the proposals is done a priori, and then the $\Nc$ scheme is applied within each cluster (i.e., small mixtures appear in the denominator of the weights). This method is called \emph{partial deterministic mixture} and in some examples is able a similar variance reduction as in the $\Nc$ method, while reducing drastically the number of proposal evaluations (see \cite[Fig. 1]{elvira2015efficient}). The \emph{overlapped partial deterministic mixture} method \cite{elvira2016overlapping} extends the framework to the case where the proposals can belong to more than one cluster. However, the way the proposals are clustered remains an open problem and few attempts have been done for optimizing the clustering (see \cite{elvira2016heretical} where the clusters are done after the sampling, using the information of the samples, and hence biasing the estimators).

When the selection of the proposals is also random, unlike in the sampling in \eqref{eq_sampling_mis}, there exist options to evaluate only the proposals that have been used for sampling (scheme $\Rb$ in \cite{elvira2019generalized}) instead of using all of them in the numerator (scheme $\Rc$ in \cite{elvira2019generalized}). A recent paper explores the $\Rb$ scheme and some of its statistical properties \cite{medina2019revisiting}.  

\subsubsection{MIS with different number of samples per proposal}

Since the seminal works of \cite{Veach95,Hesterberg95} in the computer graphics community, several works have addressed the case where the number of samples (also called counts) per proposal (also called techniques) can be different (see also \cite{Owen00}  where the authors introduce control variates in MIS). In particular, the so-called \emph{balance heuristic} estimator, proposed in \cite{Veach95} and very related to the scheme $\Nc$ in Section \eqref{sec_gen_mis} has attracted attention due to its high performance. The UIS balance heuristic estimator is given by 
\begin{equation}
\widehat I^N(f) = \sum_{j=1}^M \sum_{i=1}^{n_j}  \frac{f(\x_{j,i}) \bar{\pi}(\x_{j,i})}{\sum_{k=1}^M n_k q_k(\x_{j,i})},
\label{eq_balance_heuristic}
\end{equation}
where again $\{q_m(\x)\}_{m=1}^M$ is the set of available proposals, $\{n_m\}_{m=1}^M$ is the number of samples associated to each proposal, $N = \sum_{k=1}^{M}n_k$ is the total number of samples, and $\x_{j,i} \sim q_j(\x)$, for $i=1,...,n_j$, and for $j=1,...,M$. Regarding the denominator in  \eqref{eq_balance_heuristic}, it can be interpreted that the $N$ samples are simulated from the mixture $\sum_{k=1}^M n_k q_k(\x)$ via stratified sampling (a similar interpretation can be done in the aforementioned $\Nc$ scheme). In \cite{sbert2018multiple}, this estimator is re-visited and novel bounds are obtained. In  \cite{sbert2019generalizing}, the balance heuristic estimator of Eq. \eqref{eq_balance_heuristic} is generalized, introducing more degrees of freedom that detach the sampling and the denominator of the importance weights, being able to obtain unbiased estimators that reduce the variance with respect to the standard balance heuristic. In \cite{he2014optimal}, control variates are introduced in an IS scheme with a mixture proposal (similarly to \cite{Owen00}), and all parameters (including the mixture weights) are optimized to minimize the variance of the UIS estimator (which is jointly convex w.r.t. the mixture probabilities and the control variate regression coefficients). {More works with a variable number of samples per proposal (either fixed or optimized) include \cite{sbert2016variance,sbert2017adaptive,sbert2019optimal}.}

\subsection{Rare event estimation}
Importance sampling is often considered as a variance reduction technique, not only in the case when sampling from $\widetilde \pi$ is not possible, but also when it is possible but not efficient. A classical example is the case of Eq. \eqref{eq_integral} when $f(\x) = \mathbb{I}_{\mathcal{S}}$, where $\mathbb{I}$ is the indicator function taking value 1 for all $\x\in\mathcal{S}$, and 0 otherwise. In rare event estimation, $\mathcal{S}$ is usually a set where the target $\widetilde \pi$ has few probability mass, and hence $I$ is a small positive number. It is then not practical to simulate from the target, since most of the samples will not contribute to the estimator due to their evaluation in $\mathbb{I}_{\mathcal{S}}(\x)$ being zero. IS allows for sampling from a different distribution that will increase the efficiency of the method when $q(\x)$ is close to $\mathbb{I}_{\mathcal{S}}$. A recent MIS method called ALOE (``At least one sample'') is able to simulate from a mixture of proposals ensuring that all of them are in the integration region $\mathcal{S}$ in the case where $\widetilde \pi(\x)$ is Gaussian and $\mathcal{S}$ is the union of half-spaces defined by a set of hyperplanes (linear constraints) \cite{owen2019importance}. As an example, the authors show successful results in a problem with $5772$ constraints, in a $326$-dimensional problem with a probability of $I \approx 10^{-22}$, with just $N=10^4$ samples. ALOE has been recently applied for characterizing wireless communications systems through the estimation of the symbol error rate \cite{elvira2019efficient,elvira2019multiple,elvira2021multiple}.

\subsection{Compressed and distributed IS}
In the last years, several works have {focused on} alleviating the computational complexity, communication, or storage in intensive IS methods. This computational burden appears often when the inferential problem is challenging and requires a {large} amount of simulated samples. {This can happen because the adaptive schemes} may require many iterations, because of the high-dimensional nature of the tackled problem, and/or because a high precision (low variance) {is required} in the estimate. In \cite{martino2018group}, several compressing schemes are proposed and theoretically analyzed for assigned importance weights to groups of samples for distributed or decentralized Bayesian inference. The framework is extended in \cite{martino2018compressed,martino2021compressedPF,martino2021compressed}, where a stronger theoretical support is given, and new deterministic and random rules for compression are given. The approach in \cite{koppel2019nearly,bedi2019compressed} considers the case of a single node that keeps simulating samples and assigning them an importance weight. The bottleneck here is the storage of the samples so one needs to decide at each time if the sample is stored or discarded. A compression algorithm is introduced for building a dictionary based on greedy subspace projections and a kernel density estimator of the targeted distribution with a limited number of samples. It is shown that asymptotic bias of this method is a tunable constant depending on the kernel bandwidth parameter and a compression parameter. {Finally, some works have studied the combination of IS estimators in the distributed setting. For instance, in \cite[Section 4]{Douc07b}, independent estimators are linearly combined with the combination weights being the inverse of the variance of each estimator.  A similar approach is followed in \cite{nguyen2014improving}, using the $\ESS$ instead of the variance of the estimator (which is unknown in most practical problems). A Bayesian combination of Monte Carlo estimators is considered in \cite{luengo2015bias,luengo2018efficient}.
Note that the MIS approach is, due to its own nature, an implicit linear combination of multiple estimators (each of them using samples from one or several proposals). This perspective is exploited for instance in \cite{havran2014optimal,sbert2018multiple_b}.}

\section{Adaptive importance sampling (AIS)}
 \label{sec_ais}

Since choosing a good proposal (or set of proposals) in advance is in general impossible, a common approach is the use of adaptive importance sampling (AIS) \cite{bugallo2017adaptive}. AIS algorithms are iterative methods for a gradual learning of one or multiple proposals that aim at approximating the target pdf. Algorithm \ref{alg: standard_AIS_alg} describes a generic AIS algorithm through three basic steps: the simulation of samples from a one or several proposals  (sampling), the computation of the importance weight of each sample (weighting), and the update of the parameters that characterize the proposal(s) for repeating the previous steps in the next iteration (adaptation).  
 
 Most existing algorithms can be described in this framework that we describe with more detail. The generic AIS algorithm initializes $N$ proposals $\{q_n(\x|\thetavec_{n,1}) \}_{n=1}^N$, parametrized each of them by a vector $\thetavec_{n,1}$. Then, $K$ samples are simulated from each proposals, {$\x_{n,1}^{(k)},\;\; n=1,\ldots,N, k=1,\ldots,K$, and weighted properly. Here again many ways of sampling and weighting are possible, as it is described in Section \ref{sec_gen_mis}. At the end of the weighting step, it is possible to approximate the integral of Eq. \eqref{eq_integral} with either UIS and SNIS, and the target distribution with a discrete random measure, by using the set of weighted samples $\{\x_{n,1}^{(k)},{w}_{n,1}^{(k)}\},\;\; n=1,\ldots,N, k=1,\ldots,K$. Finally, the parameters of the $n$-th proposals are updated from $\thetavec_{n,1}$ to $\thetavec_{n,2}$. {This three-step process} is repeated until an iteration stoppage criterion is met (e.g., a maximum number of iterations, $J$, is reached). Note that at the end, {the estimators can either use} all weighted samples from iterations $1$ to $J$, or only the samples from the last iteration. 

   \begin{algorithm}[H]
    \caption{Generic AIS algorithm}
    \label{alg: standard_AIS_alg}
    {\small
      \begin{algorithmic}[1]
        \STATE \textbf{Input:} Choose $K$, $N$, $J$, and  $\{ \thetavec_{n,1}  \}_{n=1}^N$
        \FOR{$i=1,\ldots,T$}
\STATE \textbf{Sampling:}  Draw {$K$} samples from each of the $N$ proposal pdfs, $\{ q_{n,j}(\x|\thetavec_{n,j}) \}_{n=1}^N$, $\x_{n,j}^{(k)},   k=1,\ldots,K,\; n=1,\ldots,N$

\STATE \textbf{Weighting:}  Calculate the weights, $w_{n,j}^{(k)},$ for each of the generated $KN$ samples.

\STATE \textbf{Adaptation:}  Update the proposal parameters $\{\thetavec_{n,j} \}_{n=1}^N \longrightarrow \{\thetavec_{n,j+1} \}_{n=1}^N$.

        \ENDFOR
                \STATE \textbf{Output:} Return the $KNJ$ pairs $\{\x_{n,j}^{(k)},w_{n,j}^{(k)}\}$ for all $k=1,\ldots,K$, $n=1,\ldots,N$, $j=1,\ldots,J$.

      \end{algorithmic}
      }
    \end{algorithm}

The literature is vast in AIS methods and a detailed description of all of them goes beyond the scope of this paper (see \cite{bugallo2017adaptive} for a thorough review). Most of the AIS algorithms can be classified within three categories, depending on how the proposals are adapted.  Figure \ref{figure_adaptive_families} shows graphically the three families of AIS algorithms, describing the dependencies for the adaptation of the proposal parameters and the simulation of the samples. Each subplot corresponds to each family, whose description and corresponding AIS algorithms of the literature are given below.
\begin{itemize}
	\item[a)] The proposal parameters are adapted using the last set of drawn samples (e.g., {standard} PMC \cite{Cappe04}, DM-PMC \cite{elvira2017improving,elvira2017population}, N-PMC \cite{koblents2015population}, M-PMC \cite{Cappe08}, APIS \cite{martino2014adaptive,APIS15}).
	\item[b)] The proposal parameters are adapted using all drawn samples up to the latest iteration (e.g., AMIS \cite{CORNUET12}, CAIS \cite{el2018robust}, Daisee \cite{lu2018exploration}, EAMIS \cite{el2019efficient}, RS-CAIS \cite{el2019recursive}).
	\item[c)] The proposal parameters are adapted using an independent process from the samples (LAIS \cite{martino2017layered,martino2017anti,martino2015mcmc,martino2015interacting}, GAPIS \cite{elvira2015gradient}, GIS \cite{schuster2015gradient}, IMIS \cite{fasiolo2018langevin}, SL-PMC \cite{elvira2019langevin,elvira2021optimized}, HAIS \cite{mousavi2021hamiltonian}).
\end{itemize}
{}

\begin{figure*}[h!]
\centering
\subfigure[The proposal parameters are adapted using the last set of drawn samples (Standard PMC, DM-PMC, N-PMC, M-PMC, APIS).]{
\includegraphics[scale=0.31]{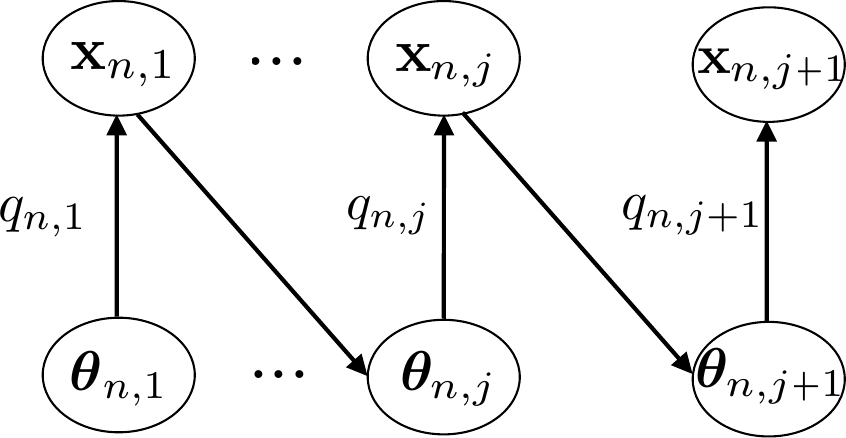}
} \quad
\subfigure[The proposal parameters are adapted using all drawn samples up to the latest iteration (AMIS, CAIS, Daisee, EAMIS, RS-CAIS).]{
\includegraphics[scale=0.31]{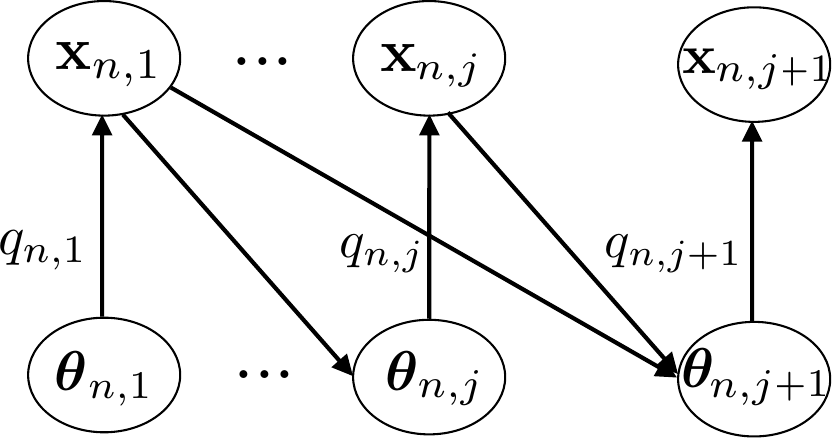}
} \quad
%\centering
\subfigure[The proposal parameters are adapted using an independent process from the samples (LAIS, GAPIS, GIS, IMIS, SL-PMC).]{
\includegraphics[scale=0.31]{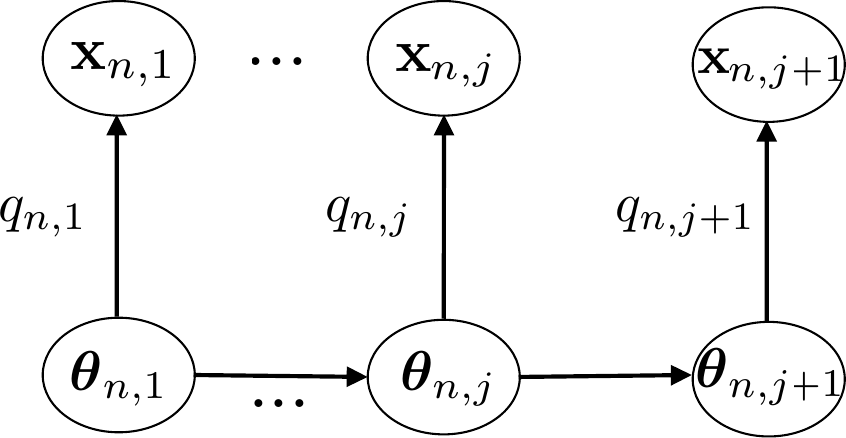}
}
\caption{{Graphical description of three possible dependencies between the adaptation of the proposal parameters $\thetavec_{n,t}$ and the samples. Note that $q_{n,t} \equiv q_{n,t}(\x|\thetavec_{n,t})$.}}
\label{figure_adaptive_families}
\end{figure*}

 In Table \ref{table_alternative}, we describe some relevant AIS algorithms according to different features: the number of proposals; the weighting scheme (\emph{nonlinear} corresponds to the clipping strategies of Section \ref{sec_crop}, \emph{standard} is equivalent to Option 1 in Section \ref{sec_gen_mis}, \emph{spatial mixture} corresponds to Option 2 with $\psi(\x) = \sum_{i=1}^N q_{i,j}(\x|\thetavec_{i,j})$, \emph{temporal mixture} corresponds to {Option 2 with} $\psi(\x) = \sum_{\tau=1}^j q_{n,\tau}(\x|\thetavec_{n,\tau})$); and the parameters that are adapted (either location and scale, or only location). In Table \ref{table_computational}, we describe the computational complexity of the same algorithms according to number of target evaluations, proposal evaluations, target evaluations per proposal, and proposal evaluations per proposal. In some AIS algorithms, the proposals converge with the number of iterations $J$, although proving this convergence {(and the associated convergence rates)} is in general a tough problem (see a recent result in \cite{akyildiz2019convergence}). For many other AIS algorithms (e.g., DM-PMC, LAIS, APIS), the proposals do not converge to any {limiting} distribution. {Converge rates have been established only for simple classes of AIS algorithms which are based on optimized parametric proposals \cite{akyildiz2019convergence}. 
All algorithms adapt the location parameters, while only few of them allow also to adapt the scale parameter.
The estimation covariance matrices is in general a complicated problem, and strategies to approximate the matrix via simpler representations (e.g., diagonal or isotropic \cite{fernandez2015probabilistic}) are exploited in CAIS \cite{el2018robust}. 
 Note that AIS-based algorithms have been also used for optimization purposes \cite{del2006sequential,akyildiz2017adaptive}.}

%%%%%%%%%%%%%
%%%%%%%%%%%%
\begin{table}[!h]
\begin{center}
\captionof{table}{{Comparison of various AIS algorithms according to different features.}}
{\footnotesize
\begin{tabular}{|c|c|c|c|c|}
\hline 
\bf{{Algorithm}} & \bf{\# proposals} & \bf{Weighting} & {\bf{Adaptation strategy}} & \bf{Parameters {adapted}}  \\
\hline\hline
Standard PMC & $N>1$ & standard & resampling & location\\
\hline
M-PMC & $N>1$ & spatial mixture & resampling & location\\
\hline
N-PMC & either & nonlinear & {moment} estimation & {location/scale}\\
\hline
LAIS & $N>1$ & {generic mixture} & {MCMC} & location\\
\hline
DM-PMC & $N>1$ & spatial mixture  & resampling & location\\
\hline
AMIS & $N=1$ & temporal mixture& moment estimation & {location/scale}\\
\hline
GAPIS & $N>1$ & spatial mixture & {gradient process} & {location/scale}\\
\hline
APIS & $N>1$ & spatial mixture & moment estimation & location \\
\hline
\end{tabular}}
\label{table_alternative}
\end{center}
\end{table}

%%%%%%%%%%%%

\begin{table}[!h]
\centering
\begin{center}
\captionof{table}{{Comparison of various AIS algorithms according to the computational complexity.}}
{\footnotesize
\begin{tabular}{|c|c|c|c|c|}
\hline 
\bf{{Algorithm}} & \bf{\# target eval} & \bf{\# proposal eval}   & \bf{\# target eval/sample} & \bf{\# proposal eval/sample} \\
\hline\hline
Standard PMC & $NJ$ & $NJ$ & $1$ & $1$ \\
\hline
N-PMC & $NJ$ & $NJ$ & $1$ & $1$  \\
\hline
M-PMC & $KJ$ & $KNJ$ & $1$ & $N$ \\
\hline
LAIS & $K(N+1)J$ & $KN^2J$ & $1+1/N$ & $N$ \\
\hline
DM-PMC & $KNJ$ & $KN^2J$ & $1$ & $N$ \\
\hline
AMIS & $KJ$ & $KJ^2$ & $1$ & $J$ \\
\hline
GAPIS & $KNJ$ & $KN^2J$ & $1$ & $N$ \\
\hline
APIS & $KNJ$ & $KN^2J$ & $1$ & $N$\\
\hline
\end{tabular}
}
\label{table_computational}
\end{center}
\end{table}

 \section*{Acknowledgements}
V.E. acknowledges support from the \emph{Agence Nationale de la Recherche} of France under PISCES project (ANR-17-CE40-0031-01).

 \bibliographystyle{abbrvnat} 
%\bibliography{bibliografia}

\begin{thebibliography}{115}
\providecommand{\natexlab}[1]{#1}
\providecommand{\url}[1]{\texttt{#1}}
\expandafter\ifx\csname urlstyle\endcsname\relax
  \providecommand{\doi}[1]{doi: #1}\else
  \providecommand{\doi}{doi: \begingroup \urlstyle{rm}\Url}\fi

\bibitem[Acton(1990)]{Acton90}
F.~S. Acton.
\newblock \emph{Numerical Methods That Work}.
\newblock The Mathematical Association of America, Washington, DC, 1990.

\bibitem[Agapiou et~al.(2017)Agapiou, Papaspiliopoulos, Sanz-Alonso, Stuart,
  et~al.]{agapiou2017importance}
S.~Agapiou, O.~Papaspiliopoulos, D.~Sanz-Alonso, A.~Stuart, et~al.
\newblock Importance sampling: Intrinsic dimension and computational cost.
\newblock \emph{Statistical Science}, 32\penalty0 (3):\penalty0 405--431, 2017.

\bibitem[Akyildiz and M{\'\i}guez(2019)]{akyildiz2019convergence}
{\"O}.~D. Akyildiz and J.~M{\'\i}guez.
\newblock Convergence rates for optimised adaptive importance samplers.
\newblock \emph{arXiv preprint arXiv:1903.12044}, 2019.

\bibitem[Akyildiz et~al.(2017)Akyildiz, Marino, and
  M{\'\i}guez]{akyildiz2017adaptive}
O.~D. Akyildiz, I.~P. Marino, and J.~M{\'\i}guez.
\newblock Adaptive noisy importance sampling for stochastic optimization.
\newblock In \emph{2017 IEEE 7th International Workshop on Computational
  Advances in Multi-Sensor Adaptive Processing (CAMSAP)}, pages 1--5. IEEE,
  2017.

\bibitem[Aus{\'i}n(2014)]{WileyAusin14_2}
M.~C. Aus{\'i}n.
\newblock Quadrature and numerical integration.
\newblock \emph{Wiley StatsRef: Statistics Reference Online (stat03875)}, pages
  1--10, 2014.

\bibitem[Bedi et~al.(2019)Bedi, Koppel, Elvira, and Sadler]{bedi2019compressed}
A.~S. Bedi, A.~Koppel, V.~Elvira, and B.~M. Sadler.
\newblock Compressed streaming importance sampling \\for efficient
  representations of localization distributions.
\newblock In \emph{2019 Asilomar Conference on Signals, Systems and Computers},
  pages 1--5. IEEE, 2019.

\bibitem[Bernardo and Smith(1994)]{Bernardo94}
J.~M. Bernardo and A.~F.~M. Smith.
\newblock \emph{{B}ayesian Theory}.
\newblock Wiley \& sons, 1994.

\bibitem[Bhadra and Ionides(2016)]{Bhadra16}
A.~Bhadra and E.~L. Ionides.
\newblock Adaptive particle allocation in iterated sequential {M}onte {C}arlo
  via approximating meta-models.
\newblock \emph{Statistics and Computing}, 26\penalty0 (1-2):\penalty0
  393--407, 2016.

\bibitem[Box and Tiao(1973)]{Box73}
G.~E.~P. Box and G.~C. Tiao.
\newblock \emph{{B}ayesian Inference in Statistical Analysis}.
\newblock Wiley \& sons, 1973.

\bibitem[Branchini and Elvira(2020)]{branchini2020optimized}
N.~Branchini and V.~Elvira.
\newblock Optimized auxiliary particle filters.
\newblock \emph{arXiv preprint arXiv:2011.09317}, 2020.

\bibitem[Bugallo et~al.(2017)Bugallo, Elvira, Martino, Luengo, M\'iguez, and
  Djuric]{bugallo2017adaptive}
M.~F. Bugallo, V.~Elvira, L.~Martino, D.~Luengo, J.~M\'iguez, and P.~M. Djuric.
\newblock Adaptive importance sampling: The past, the present, and the future.
\newblock \emph{IEEE Signal Process. Mag.}, 34\penalty0 (4):\penalty0 60--79,
  2017.

\bibitem[Burden and Faires(2000)]{Burden00}
R.~L. Burden and J.~D. Faires.
\newblock \emph{Numerical Analysis}.
\newblock Brooks Cole, 2000.

\bibitem[Capp\'e et~al.(2004)Capp\'e, Guillin, Marin, and Robert]{Cappe04}
O.~Capp\'e, A.~Guillin, J.~M. Marin, and C.~P. Robert.
\newblock Population {M}onte {C}arlo.
\newblock \emph{Journal of Comp. and Graphical Statistics}, 13\penalty0
  (4):\penalty0 907--929, 2004.

\bibitem[Capp\'e et~al.(2008)Capp\'e, Douc, Guillin, Marin, and
  Robert]{Cappe08}
O.~Capp\'e, R.~Douc, A.~Guillin, J.~M. Marin, and C.~P. Robert.
\newblock Adaptive importance sampling in general mixture classes.
\newblock \emph{Stat. Comput.}, 18:\penalty0 447--459, 2008.

\bibitem[Chatterjee et~al.(2018)Chatterjee, Diaconis,
  et~al.]{chatterjee2018sample}
S.~Chatterjee, P.~Diaconis, et~al.
\newblock The sample size required in importance sampling.
\newblock \emph{The Annals of Applied Probability}, 28\penalty0 (2):\penalty0
  1099--1135, 2018.

\bibitem[Christen(2014)]{christen2014g}
J.~A. Christen.
\newblock Gibbs sampling.
\newblock \emph{Wiley StatsRef: Statistics Reference Online}, pages 1--9, 2014.

\bibitem[Cornuet et~al.(2012)Cornuet, Marin, Mira, and Robert]{CORNUET12}
J.~M. Cornuet, J.~M. Marin, A.~Mira, and C.~P. Robert.
\newblock Adaptive multiple importance sampling.
\newblock \emph{Scandinavian Journal of Statistics}, 39\penalty0 (4):\penalty0
  798--812, December 2012.

\bibitem[Djuric et~al.(2007)Djuric, Lu, and Bugallo]{djuric2007multiple}
P.~M. Djuric, T.~Lu, and M.~F. Bugallo.
\newblock Multiple particle filtering.
\newblock In \emph{2007 IEEE International Conference on Acoustics, Speech and
  Signal Processing-ICASSP'07}, volume~3, pages III--1181. IEEE, 2007.

\bibitem[Douc et~al.(2005)Douc, Capp\'e, and Moulines]{Douc05}
R.~Douc, O.~Capp\'e, and E.~Moulines.
\newblock Comparison of resampling schemes for particle filtering.
\newblock In \emph{Proc. 4${}^{th}$ Int. Symp. on Image and Signal Processing
  and Analysis}, pages 64--69, September 2005.

\bibitem[Douc et~al.(2007)Douc, Guillin, Marin, and Robert]{Douc07b}
R.~Douc, A.~Guillin, J.~M. Marin, and C.~P. Robert.
\newblock Minimum variance importance sampling via population {M}onte {C}arlo.
\newblock \emph{ESAIM: Probability and Statistics}, 11:\penalty0 427--447,
  2007.

\bibitem[Doucet and Johansen(2009)]{doucet2009tutorial}
A.~Doucet and A.~M. Johansen.
\newblock A tutorial on particle filtering and smoothing: Fifteen years later.
\newblock \emph{Handbook of nonlinear filtering}, 12\penalty0
  (656-704):\penalty0 3, 2009.

\bibitem[Doucet et~al.(2000)Doucet, De~Freitas, Murphy, and
  Russell]{doucet2000rao}
A.~Doucet, N.~De~Freitas, K.~Murphy, and S.~Russell.
\newblock Rao-blackwellised particle filtering for dynamic {B}ayesian networks.
\newblock In \emph{Proceedings of the Sixteenth conference on Uncertainty in
  artificial intelligence}, pages 176--183. Morgan Kaufmann Publishers Inc.,
  2000.

\bibitem[Dunn and Shultis(2011)]{Dunn2011}
W.~L. Dunn and J.~K. Shultis.
\newblock \emph{Exploring {M}onte {C}arlo Methods}.
\newblock Elsevier Science, Amsterdam (The Netherlands), 2011.

\bibitem[El-Laham et~al.(2018)El-Laham, Elvira, and Bugallo]{el2018robust}
Y.~El-Laham, V.~Elvira, and M.~F. Bugallo.
\newblock Robust covariance adaptation in adaptive importance sampling.
\newblock \emph{IEEE Signal Processing Letters}, 25\penalty0 (7):\penalty0
  1049--1053, 2018.

\bibitem[El-Laham et~al.(2019{\natexlab{a}})El-Laham, Elvira, and
  Bugallo]{el2019recursive}
Y.~El-Laham, V.~Elvira, and M.~F. Bugallo.
\newblock Recursive shrinkage covariance learning in adaptive importance
  sampling.
\newblock In \emph{Proc. IEEE Int. Work. Comput. Adv. Multi-Sensor Adap.
  Process. (CAMSAP 2019)}, pages 1--5, 2019{\natexlab{a}}.

\bibitem[El-Laham et~al.(2019{\natexlab{b}})El-Laham, Martino, Elvira, and
  Bugallo]{el2019efficient}
Y.~El-Laham, L.~Martino, V.~Elvira, and M.~F. Bugallo.
\newblock Efficient adaptive multiple importance sampling.
\newblock In \emph{2019 27th European Signal Processing Conference (EUSIPCO)},
  pages 1--5. IEEE, 2019{\natexlab{b}}.

\bibitem[Elvira and Chouzenoux(2019)]{elvira2019langevin}
V.~Elvira and {\'E}.~Chouzenoux.
\newblock Langevin-based strategy for efficient proposal adaptation in
  population {M}onte {C}arlo.
\newblock In \emph{ICASSP 2019-2019 IEEE International Conference on Acoustics,
  Speech and Signal Processing (ICASSP)}, pages 5077--5081. IEEE, 2019.

\bibitem[Elvira and Chouzenoux(2021)]{elvira2021optimized}
V.~Elvira and E.~Chouzenoux.
\newblock Optimized population monte carlo.
\newblock 2021.

\bibitem[Elvira and Santamar{\'\i}a(2019{\natexlab{a}})]{elvira2019efficient}
V.~Elvira and I.~Santamar{\'\i}a.
\newblock Efficient ser estimation for mimo detectors via importance sampling
  schemes.
\newblock In \emph{2019 Asilomar Conference on Signals, Systems and Computers},
  pages 1--5. IEEE, 2019{\natexlab{a}}.

\bibitem[Elvira and Santamar{\'\i}a(2019{\natexlab{b}})]{elvira2019multiple}
V.~Elvira and I.~Santamar{\'\i}a.
\newblock Multiple importance sampling for efficient symbol error rate
  estimation.
\newblock \emph{IEEE Signal Processing Letters}, 26\penalty0 (3):\penalty0
  420--424, 2019{\natexlab{b}}.

\bibitem[Elvira and Santamaria(2021)]{elvira2021multiple}
V.~Elvira and I.~Santamaria.
\newblock Multiple importance sampling for symbol error rate estimation of
  maximum-likelihood detectors in mimo channels.
\newblock \emph{IEEE Transactions on Signal Processing}, 69:\penalty0
  1200--1212, 2021.

\bibitem[Elvira et~al.(2015{\natexlab{a}})Elvira, Martino, Luengo, and
  Bugallo]{elvira2015efficient}
V.~Elvira, L.~Martino, D.~Luengo, and M.~F. Bugallo.
\newblock Efficient multiple importance sampling estimators.
\newblock \emph{Signal Processing Letters, IEEE}, 22\penalty0 (10):\penalty0
  1757--1761, 2015{\natexlab{a}}.

\bibitem[Elvira et~al.(2015{\natexlab{b}})Elvira, Martino, Luengo, and
  Corander]{elvira2015gradient}
V.~Elvira, L.~Martino, L.~Luengo, and J.~Corander.
\newblock A gradient adaptive population importance sampler.
\newblock In \emph{Proc. IEEE Int. Conf. Acoust., Speech Signal Process.
  (ICASSP 2015)}, pages 4075--4079, Brisbane, Australia, 19-24 April
  2015{\natexlab{b}}.

\bibitem[Elvira et~al.(2016{\natexlab{a}})Elvira, Martino, Luengo, and
  Bugallo]{elvira2016heretical}
V.~Elvira, L.~Martino, D.~Luengo, and M.~F. Bugallo.
\newblock Heretical multiple importance sampling.
\newblock \emph{IEEE Signal Processing Letters}, 23\penalty0 (10):\penalty0
  1474--1478, 2016{\natexlab{a}}.

\bibitem[Elvira et~al.(2016{\natexlab{b}})Elvira, Martino, Luengo, and
  Bugallo]{elvira2016overlapping}
V.~Elvira, L.~Martino, D.~Luengo, and M.~F. Bugallo.
\newblock Multiple importance sampling with overlapping sets of proposals.
\newblock \emph{IEEE Workshop on Statistical Signal Processing (SSP)},
  2016{\natexlab{b}}.

\bibitem[Elvira et~al.(2017{\natexlab{a}})Elvira, Martino, Luengo, and
  Bugallo]{elvira2017improving}
V.~Elvira, L.~Martino, D.~Luengo, and M.~F. Bugallo.
\newblock Improving {P}opulation {M}onte {C}arlo: Alternative weighting and
  resampling schemes.
\newblock \emph{Sig. Process.}, 131\penalty0 (12):\penalty0 77--91,
  2017{\natexlab{a}}.

\bibitem[Elvira et~al.(2017{\natexlab{b}})Elvira, Martino, Luengo, and
  Bugallo]{elvira2017population}
V.~Elvira, L.~Martino, D.~Luengo, and M.~F. Bugallo.
\newblock Population {M}onte {C}arlo schemes with reduced path degeneracy.
\newblock In \emph{Proc. IEEE Int. Work. Comput. Adv. Multi-Sensor Adap.
  Process. (CAMSAP 2017)}, pages 1--5, 2017{\natexlab{b}}.

\bibitem[Elvira et~al.(2017{\natexlab{c}})Elvira, M{\'\i}guez, and
  Djuri{\'c}]{elvira2017adapting}
V.~Elvira, J.~M{\'\i}guez, and P.~Djuri{\'c}.
\newblock Adapting the number of particles in sequential monte carlo methods
  through an online scheme for convergence assessment.
\newblock \emph{IEEE Transactions on Signal Processing}, 65\penalty0
  (7):\penalty0 1781--1794, 2017{\natexlab{c}}.

\bibitem[Elvira et~al.(2018)Elvira, Martino, Bugallo, and
  Djuri{\'c}]{elvira2018search}
V.~Elvira, L.~Martino, M.~F. Bugallo, and P.~M. Djuri{\'c}.
\newblock In search for improved auxiliary particle filters.
\newblock In \emph{2018 26th European Signal Processing Conference (EUSIPCO)},
  pages 1637--1641. IEEE, 2018.

\bibitem[Elvira et~al.(2019{\natexlab{a}})Elvira, Closas, and
  Martino]{elvira2019gauss}
V.~Elvira, P.~Closas, and L.~Martino.
\newblock {G}auss-{H}ermite quadrature for non-gaussian inference via an
  importance sampling interpretation.
\newblock In \emph{2019 27th European Signal Processing Conference (EUSIPCO)},
  pages 1--5. IEEE, 2019{\natexlab{a}}.

\bibitem[Elvira et~al.(2019{\natexlab{b}})Elvira, Martino, Bugallo, and
  Djuric]{elvira2019elucidating}
V.~Elvira, L.~Martino, M.~F. Bugallo, and P.~M. Djuric.
\newblock Elucidating the auxiliary particle filter via multiple importance
  sampling [lecture notes].
\newblock \emph{IEEE Signal Processing Magazine}, 36\penalty0 (6):\penalty0
  145--152, 2019{\natexlab{b}}.

\bibitem[Elvira et~al.(2019{\natexlab{c}})Elvira, Martino, Luengo, and
  Bugallo]{elvira2019generalized}
V.~Elvira, L.~Martino, D.~Luengo, and M.~F. Bugallo.
\newblock Generalized multiple importance sampling.
\newblock \emph{Statistical Science}, 34\penalty0 (1):\penalty0 129--155,
  2019{\natexlab{c}}.

\bibitem[Elvira et~al.(2019{\natexlab{d}})Elvira, M{\'\i}guez, and
  Djuri{\'c}]{elvira2019new}
V.~Elvira, J.~M{\'\i}guez, and P.~M. Djuri{\'c}.
\newblock New results on particle filters with adaptive number of particles.
\newblock \emph{arXiv preprint arXiv:1911.01383}, 2019{\natexlab{d}}.

\bibitem[Elvira et~al.(2020)Elvira, Martino, and Closas]{elvira2020importance}
V.~Elvira, L.~Martino, and P.~Closas.
\newblock Importance gaussian quadrature.
\newblock \emph{IEEE Transactions on Signal Processing}, 69:\penalty0 474--488,
  2020.

\bibitem[Elvira et~al.(2022)Elvira, Martino, and Robert]{elvira2018rethinking}
V.~Elvira, L.~Martino, and C.~P. Robert.
\newblock Rethinking the effective sample size.
\newblock \emph{International Statistical Review, to appear in}, 2022.

\bibitem[Fasiolo et~al.(2018)Fasiolo, de~Melo, and
  Maskell]{fasiolo2018langevin}
M.~Fasiolo, F.~E. de~Melo, and S.~Maskell.
\newblock Langevin incremental mixture importance sampling.
\newblock \emph{Stat. Comput.}, 28\penalty0 (3):\penalty0 549--561, 2018.

\bibitem[Fernandez-Bes et~al.(2015)Fernandez-Bes, Elvira, and
  Van~Vaerenbergh]{fernandez2015probabilistic}
J.~Fernandez-Bes, V.~Elvira, and S.~Van~Vaerenbergh.
\newblock A probabilistic least-mean-squares filter.
\newblock In \emph{2015 IEEE International Conference on Acoustics, Speech and
  Signal Processing (ICASSP)}, pages 2199--2203. IEEE, 2015.

\bibitem[Gentle(2004)]{Gentle04}
J.~E. Gentle.
\newblock \emph{Random Number Generation and {M}onte {C}arlo Methods}.
\newblock Springer, 2004.

\bibitem[Gordon et~al.(1993)Gordon, Salmond, and Smith]{Gordon93}
N.~Gordon, D.~Salmond, and A.~F.~M. Smith.
\newblock Novel approach to nonlinear and non-{G}aussian {B}ayesian state
  estimation.
\newblock \emph{IEE Proceedings-F Radar and Signal Processing}, 140:\penalty0
  107--113, 1993.

\bibitem[Havran and Sbert(2014)]{havran2014optimal}
V.~Havran and M.~Sbert.
\newblock Optimal combination of techniques in multiple importance sampling.
\newblock In \emph{Proceedings of the 13th ACM SIGGRAPH International
  Conference on Virtual-Reality Continuum and its Applications in Industry},
  pages 141--150, 2014.

\bibitem[He and Owen(2014)]{he2014optimal}
H.~Y. He and A.~B. Owen.
\newblock Optimal mixture weights in multiple importance sampling.
\newblock \emph{arXiv preprint arXiv:1411.3954}, 2014.

\bibitem[Hesterberg(1995)]{Hesterberg95}
T.~Hesterberg.
\newblock Weighted average importance sampling and defensive mixture
  distributions.
\newblock \emph{Technometrics}, 37\penalty0 (2):\penalty0 185--194, 1995.

\bibitem[Huggins et~al.(2019)Huggins, Roy, et~al.]{huggins2019sequential}
J.~H. Huggins, D.~M. Roy, et~al.
\newblock Sequential {M}onte {C}arlo as approximate sampling: bounds, adaptive
  resampling via $infty$-ess, and an application to particle gibbs.
\newblock \emph{Bernoulli}, 25\penalty0 (1):\penalty0 584--622, 2019.

\bibitem[Ionides(2008)]{ionides2008truncated}
E.~L. Ionides.
\newblock Truncated importance sampling.
\newblock \emph{Journal of Computational and Graphical Statistics}, 17\penalty0
  (2):\penalty0 295--311, 2008.

\bibitem[Jaeckel(2002)]{Jaeckel02}
P.~Jaeckel.
\newblock \emph{{M}onte {C}arlo Methods in Finance}.
\newblock Wiley, 2002.

\bibitem[Kahn(1950)]{kahn1950random}
H.~Kahn.
\newblock Random sampling ({M}onte {C}arlo) techniques in neutron attenuation
  problems.
\newblock \emph{Nucleonics}, 6\penalty0 (5):\penalty0 27--passim, 1950.

\bibitem[Koblents and M{\'\i}guez(2015)]{koblents2015population}
E.~Koblents and J.~M{\'\i}guez.
\newblock A population {M}onte {C}arlo scheme with transformed weights and its
  application to stochastic kinetic models.
\newblock \emph{Statistics and Computing}, 25\penalty0 (2):\penalty0 407--425,
  2015.

\bibitem[Kong(1992)]{Kong92}
A.~Kong.
\newblock A note on importance sampling using standardized weights.
\newblock \emph{University of Chicago, Dept. of Statistics, Tech. Rep}, 348,
  1992.

\bibitem[Kong(2014)]{kong2014importance}
A.~Kong.
\newblock Importance sampling.
\newblock \emph{Wiley StatsRef: Statistics Reference Online}, 2014.

\bibitem[Koppel et~al.(2021)Koppel, Bedi, Sadler, and Elvira]{koppel2019nearly}
A.~Koppel, A.~S. Bedi, B.~M. Sadler, and V.~Elvira.
\newblock Nearly consistent finite particle estimates in streaming importance
  sampling.
\newblock \emph{IEEE Transactions on Signal Processing}, 69:\penalty0
  6401--6415, 2021.

\bibitem[Kotecha and Djuri\'c(2003)]{Kotecha03a}
J.~Kotecha and P.~M. Djuri\'c.
\newblock {G}aussian particle filtering.
\newblock \emph{IEEE Transactions Signal Processing}, 51\penalty0
  (10):\penalty0 2592--2601, October 2003.

\bibitem[Kroese et~al.(2011)Kroese, Taimre, and Botev]{Kroese11}
D.~Kroese, T.~Taimre, and Z.~Botev.
\newblock \emph{Handbook of {M}onte {C}arlo Methods}.
\newblock Wiley Series in Probability and Statistics, John Wiley and Sons, New
  York, 2011.

\bibitem[Kythe and Schaferkotter(2004)]{Kythe04}
P.~K. Kythe and M.~R. Schaferkotter.
\newblock \emph{Handbook of Computational Methods for Integration}.
\newblock Chapman and Hall/CRC, 2004.

\bibitem[Lee and Whiteley(2015)]{Lee15}
A.~Lee and N.~Whiteley.
\newblock Variance estimation and allocation in the particle filter.
\newblock \emph{arXiv:1509.00394v1 [stat.CO]}, 2015.

\bibitem[Lee(2014)]{WileyLee14}
P.~M. Lee.
\newblock {B}ayesian inference.
\newblock \emph{Wiley StatsRef: Statistics Reference Online (stat00207.pub2)},
  pages 1--9, 2014.

\bibitem[Li et~al.(2015)Li, Bolic, and Djuric]{li2015resampling}
T.~Li, M.~Bolic, and P.~M. Djuric.
\newblock Resampling methods for particle filtering: Classification,
  implementation, and strategies.
\newblock \emph{IEEE Signal Processing Magazine}, 32\penalty0 (3):\penalty0
  70--86, 2015.

\bibitem[Liu(2004)]{Liu04b}
J.~S. Liu.
\newblock \emph{{M}onte {C}arlo Strategies in Scientific Computing}.
\newblock Springer, 2004.

\bibitem[Lu et~al.(2018)Lu, Rainforth, Zhou, van~de Meent, and
  Teh]{lu2018exploration}
X.~Lu, T.~Rainforth, Y.~Zhou, J.-W. van~de Meent, and Y.~W. Teh.
\newblock On exploration, exploitation and learning in adaptive importance
  sampling.
\newblock \emph{arXiv preprint arXiv:1810.13296}, 2018.

\bibitem[Luengo et~al.(2015)Luengo, Martino, Elvira, and
  Bugallo]{luengo2015bias}
D.~Luengo, L.~Martino, V.~Elvira, and M.~Bugallo.
\newblock Bias correction for distributed bayesian estimators.
\newblock In \emph{2015 IEEE 6th International Workshop on Computational
  Advances in Multi-Sensor Adaptive Processing (CAMSAP)}, pages 253--256. IEEE,
  2015.

\bibitem[Luengo et~al.(2018)Luengo, Martino, Elvira, and
  Bugallo]{luengo2018efficient}
D.~Luengo, L.~Martino, V.~Elvira, and M.~Bugallo.
\newblock Efficient linear fusion of partial estimators.
\newblock \emph{Digital Signal Processing}, 78:\penalty0 265--283, 2018.

\bibitem[Luengo et~al.(2020)Luengo, Martino, Bugallo, Elvira, and
  S{\"a}rkk{\"a}]{luengo2020survey}
D.~Luengo, L.~Martino, M.~Bugallo, V.~Elvira, and S.~S{\"a}rkk{\"a}.
\newblock A survey of monte carlo methods for parameter estimation.
\newblock \emph{EURASIP Journal on Advances in Signal Processing},
  2020:\penalty0 1--62, 2020.

\bibitem[Martino and Elvira(2017)]{martino2014metropolis}
L.~Martino and V.~Elvira.
\newblock Metropolis sampling.
\newblock \emph{Wiley StatsRef: Statistics Reference Online}, pages 1--18,
  2017.

\bibitem[Martino and Elvira(2018)]{martino2018compressed}
L.~Martino and V.~Elvira.
\newblock Compressed {M}onte {C}arlo for distributed {B}ayesian inference.
\newblock \emph{viXra:1811.0505}, 2018.

\bibitem[Martino and Elvira(2021)]{martino2021compressedPF}
L.~Martino and V.~Elvira.
\newblock Compressed monte carlo with application in particle filtering.
\newblock \emph{Information Sciences}, 553:\penalty0 331--352, 2021.

\bibitem[Martino et~al.(2014)Martino, Elvira, Luengo, and
  Corander]{martino2014adaptive}
L.~Martino, V.~Elvira, D.~Luengo, and J.~Corander.
\newblock An adaptive population importance sampler.
\newblock In \emph{Acoustics, Speech and Signal Processing (ICASSP), 2014 IEEE
  International Conference on}, pages 8038--8042. IEEE, 2014.

\bibitem[Martino et~al.(2015{\natexlab{a}})Martino, Elvira, Luengo, and
  Corander]{APIS15}
L.~Martino, V.~Elvira, D.~Luengo, and J.~Corander.
\newblock An adaptive population importance sampler: Learning from the
  uncertanity.
\newblock \emph{IEEE Transactions on Signal Processing}, 63\penalty0
  (16):\penalty0 4422--4437, 2015{\natexlab{a}}.

\bibitem[Martino et~al.(2015{\natexlab{b}})Martino, Elvira, Luengo, and
  Corander]{martino2015interacting}
L.~Martino, V.~Elvira, D.~Luengo, and J.~Corander.
\newblock Interacting parallel markov adaptive importance sampling.
\newblock In \emph{European Signal Processing Conference (EUSIPCO)}, pages
  1--5, 2015{\natexlab{b}}.

\bibitem[Martino et~al.(2015{\natexlab{c}})Martino, Elvira, Luengo, and
  Corander]{martino2015mcmc}
L.~Martino, V.~Elvira, D.~Luengo, and J.~Corander.
\newblock Mcmc-driven adaptive multiple importance sampling.
\newblock In \emph{Interdisciplinary Bayesian Statistics}, pages 97--109.
  Springer, 2015{\natexlab{c}}.

\bibitem[Martino et~al.(2016{\natexlab{a}})Martino, Elvira, and
  Louzada]{martino2016alternative}
L.~Martino, V.~Elvira, and F.~Louzada.
\newblock Alternative effective sample size measures for importance sampling.
\newblock In \emph{2016 IEEE Statistical Signal Processing Workshop (SSP)},
  pages 1--5. IEEE, 2016{\natexlab{a}}.

\bibitem[Martino et~al.(2016{\natexlab{b}})Martino, Elvira, and
  Louzada]{martino2016weighting}
L.~Martino, V.~Elvira, and F.~Louzada.
\newblock Weighting a resampled particle in sequential {M}onte {C}arlo.
\newblock In \emph{2016 IEEE Statistical Signal Processing Workshop (SSP)},
  pages 1--5. IEEE, 2016{\natexlab{b}}.

\bibitem[Martino et~al.(2017{\natexlab{a}})Martino, Elvira, and
  Louzada]{martino2017effective}
L.~Martino, V.~Elvira, and F.~Louzada.
\newblock Effective sample size for importance sampling based on discrepancy
  measures.
\newblock \emph{Signal Processing}, 131:\penalty0 386--401, 2017{\natexlab{a}}.

\bibitem[Martino et~al.(2017{\natexlab{b}})Martino, Elvira, and
  Luengo]{martino2017anti}
L.~Martino, V.~Elvira, and D.~Luengo.
\newblock Anti-tempered layered adaptive importance sampling.
\newblock In \emph{2017 22nd International Conference on Digital Signal
  Processing (DSP)}, pages 1--5. IEEE, 2017{\natexlab{b}}.

\bibitem[Martino et~al.(2017{\natexlab{c}})Martino, Elvira, Luengo, and
  Corander]{martino2017layered}
L.~Martino, V.~Elvira, D.~Luengo, and J.~Corander.
\newblock Layered adaptive importance sampling.
\newblock \emph{Statistics and Computing}, 27\penalty0 (3):\penalty0 599--623,
  2017{\natexlab{c}}.

\bibitem[Martino et~al.(2018{\natexlab{a}})Martino, Elvira, and
  Camps-Valls]{martino2018group}
L.~Martino, V.~Elvira, and G.~Camps-Valls.
\newblock Group importance sampling for particle filtering and mcmc.
\newblock \emph{Digital Signal Processing}, 82:\penalty0 133--151,
  2018{\natexlab{a}}.

\bibitem[Martino et~al.(2018{\natexlab{b}})Martino, Elvira, M\'iguez,
  Art{\'e}s-Rodr\'iguez, and Djuri{\'c}]{martino2018comparison}
L.~Martino, V.~Elvira, J.~M\'iguez, A.~Art{\'e}s-Rodr\'iguez, and
  P.~Djuri{\'c}.
\newblock A comparison of clipping strategies for importance sampling.
\newblock In \emph{2018 IEEE Statistical Signal Processing Workshop (SSP)},
  pages 558--562. IEEE, 2018{\natexlab{b}}.

\bibitem[Martino et~al.(2021)Martino, Elvira, L{\'o}pez-Santiago, and
  Camps-Valls]{martino2021compressed}
L.~Martino, V.~Elvira, J.~L{\'o}pez-Santiago, and G.~Camps-Valls.
\newblock Compressed particle methods for expensive models with application in
  astronomy and remote sensing.
\newblock \emph{IEEE Transactions on Aerospace and Electronic Systems}, 2021.

\bibitem[Medina-Aguayo and Everitt(2019)]{medina2019revisiting}
F.~J. Medina-Aguayo and R.~G. Everitt.
\newblock Revisiting the balance heuristic for estimating normalising
  constants.
\newblock \emph{arXiv preprint arXiv:1908.06514}, 2019.

\bibitem[Medova(2015)]{WileyMedova08}
E.~Medova.
\newblock {B}ayesian {A}nalysis and {M}arkov {C}hain {M}onte {C}arlo
  simulation.
\newblock \emph{Wiley StatsRef: Statistics Reference Online (stat03616)}, pages
  1--12, 2015.

\bibitem[M{\'\i}guez(2017)]{miguez2017performance}
J.~M{\'\i}guez.
\newblock On the performance of nonlinear importance samplers and population
  {M}onte {C}arlo schemes.
\newblock In \emph{2017 22nd International Conference on Digital Signal
  Processing (DSP)}, pages 1--5. IEEE, 2017.

\bibitem[Miguez et~al.(2018)Miguez, Mari{\~n}o, and
  V{\'a}zquez]{miguez2018analysis}
J.~Miguez, I.~P. Mari{\~n}o, and M.~A. V{\'a}zquez.
\newblock Analysis of a nonlinear importance sampling scheme for {B}ayesian
  parameter estimation in state-space models.
\newblock \emph{Signal Processing}, 142:\penalty0 281--291, 2018.

\bibitem[Moral et~al.(2006)Moral, Doucet, and Jasra]{del2006sequential}
P.~D. Moral, A.~Doucet, and A.~Jasra.
\newblock Sequential {M}onte {C}arlo samplers.
\newblock \emph{J. R. Stat. Soc. Ser. B Stat. Methodol.}, 68\penalty0
  (3):\penalty0 411--436, 2006.

\bibitem[Mousavi et~al.(2021)Mousavi, Monsefi, and
  Elvira]{mousavi2021hamiltonian}
A.~Mousavi, R.~Monsefi, and V.~Elvira.
\newblock Hamiltonian adaptive importance sampling.
\newblock \emph{IEEE Signal Processing Letters}, 2021.

\bibitem[Nguyen et~al.(2014)Nguyen, Septier, Peters, and
  Delignon]{nguyen2014improving}
T.~L.~T. Nguyen, F.~Septier, G.~W. Peters, and Y.~Delignon.
\newblock Improving smc sampler estimate by recycling all past simulated
  particles.
\newblock In \emph{Statistical Signal Processing (SSP), 2014 IEEE Workshop on},
  pages 117--120. IEEE, 2014.

\bibitem[Owen and Zhou(2000)]{Owen00}
A.~Owen and Y.~Zhou.
\newblock Safe and effective importance sampling.
\newblock \emph{Journal of the American Statistical Association}, 95\penalty0
  (449):\penalty0 135--143, 2000.

\bibitem[Owen(2013)]{owen2013monte}
A.~B. Owen.
\newblock \emph{{M}onte {C}arlo theory, methods and examples}.
\newblock 2013.

\bibitem[Owen et~al.(2019)Owen, Maximov, Chertkov, et~al.]{owen2019importance}
A.~B. Owen, Y.~Maximov, M.~Chertkov, et~al.
\newblock Importance sampling the union of rare events with an application to
  power systems analysis.
\newblock \emph{Electronic Journal of Statistics}, 13\penalty0 (1):\penalty0
  231--254, 2019.

\bibitem[Pitt and Shephard(2001)]{Pitt01}
M.~K. Pitt and N.~Shephard.
\newblock Auxiliary variable based particle filters.
\newblock In A.~Doucet, N.~de~Freitas, and N.~Gordon, editors, \emph{Sequential
  {M}onte {C}arlo Methods in Practice}, chapter~13, pages 273--293. Springer,
  2001.

\bibitem[Plybon(1992)]{Plybon92}
B.~F. Plybon.
\newblock \emph{An Introduction to Applied Numerical Analysis}.
\newblock PWS-Kent, Boston, MA, 1992.

\bibitem[Robert(2007)]{Robert07}
C.~P. Robert.
\newblock \emph{The {B}ayesian Choice}.
\newblock Springer, 2007.

\bibitem[Robert(2014)]{robert2014m}
C.~P. Robert.
\newblock {M}onte {C}arlo methods.
\newblock \emph{Wiley StatsRef: Statistics Reference Online}, pages 1--13,
  2014.

\bibitem[Robert(2015)]{robert2015}
C.~P. Robert.
\newblock The metropolis–hastings algorithm.
\newblock \emph{Wiley StatsRef: Statistics Reference Online}, pages 1--13,
  2015.

\bibitem[Robert and Casella(2004)]{Robert04}
C.~P. Robert and G.~Casella.
\newblock \emph{{M}onte {C}arlo Statistical Methods}.
\newblock Springer, 2004.

\bibitem[Ryu and Boyd(2014)]{ryu2014adaptive}
E.~K. Ryu and S.~P. Boyd.
\newblock Adaptive importance sampling via stochastic convex programming.
\newblock \emph{arXiv preprint arXiv:1412.4845}, 2014.

\bibitem[Sbert and Elvira(2022)]{sbert2019generalizing}
M.~Sbert and V.~Elvira.
\newblock Generalizing the balance heuristic estimator in multiple importance
  sampling.
\newblock \emph{Entropy}, 24\penalty0 (2):\penalty0 191, 2022.

\bibitem[Sbert and Havran(2017)]{sbert2017adaptive}
M.~Sbert and V.~Havran.
\newblock Adaptive multiple importance sampling for general functions.
\newblock \emph{The Visual Computer}, 33\penalty0 (6-8):\penalty0 845--855,
  2017.

\bibitem[Sbert et~al.(2016)Sbert, Havran, and Szirmay-Kalos]{sbert2016variance}
M.~Sbert, V.~Havran, and L.~Szirmay-Kalos.
\newblock Variance analysis of multi-sample and one-sample multiple importance
  sampling.
\newblock In \emph{Computer Graphics Forum}, volume~35, pages 451--460. Wiley
  Online Library, 2016.

\bibitem[Sbert et~al.(2018{\natexlab{a}})Sbert, Havran, and
  Szirmay-Kalos]{sbert2018multiple}
M.~Sbert, V.~Havran, and L.~Szirmay-Kalos.
\newblock Multiple importance sampling revisited: breaking the bounds.
\newblock \emph{EURASIP Journal on Advances in Signal Processing},
  2018\penalty0 (1):\penalty0 15, 2018{\natexlab{a}}.

\bibitem[Sbert et~al.(2018{\natexlab{b}})Sbert, Havran, Szirmay-Kalos, and
  Elvira]{sbert2018multiple_b}
M.~Sbert, V.~Havran, L.~Szirmay-Kalos, and V.~Elvira.
\newblock Multiple importance sampling characterization by weighted mean
  invariance.
\newblock \emph{The Visual Computer}, 34\penalty0 (6-8):\penalty0 843--852,
  2018{\natexlab{b}}.

\bibitem[Sbert et~al.(2019)Sbert, Havran, and Szirmay-Kalos]{sbert2019optimal}
M.~Sbert, V.~Havran, and L.~Szirmay-Kalos.
\newblock Optimal deterministic mixture sampling.
\newblock In \emph{Eurographics (Short Papers)}, pages 73--76, 2019.

\bibitem[Schuster(2015)]{schuster2015gradient}
I.~Schuster.
\newblock Gradient importance sampling.
\newblock Technical report, 2015.
\newblock https://arxiv.org/abs/1507.05781.

\bibitem[Taimre et~al.(2019)Taimre, Kroese, and Botev]{taimre2019monte}
T.~Taimre, D.~P. Kroese, and Z.~I. Botev.
\newblock {M}onte {C}arlo methods.
\newblock \emph{Wiley StatsRef: Statistics Reference Online DOI}, 10:\penalty0
  9781118445112, 2019.

\bibitem[Tokdar and Kass(2010)]{tokdar2010importance}
S.~T. Tokdar and R.~E. Kass.
\newblock Importance sampling: a review.
\newblock \emph{Wiley Interdisciplinary Reviews: Computational Statistics},
  2\penalty0 (1):\penalty0 54--60, 2010.

\bibitem[Veach and Guibas(1995)]{Veach95}
E.~Veach and L.~Guibas.
\newblock Optimally combining sampling techniques for {M}onte {C}arlo
  rendering.
\newblock \emph{In SIGGRAPH 1995 Proceedings}, pages 419--428, 1995.

\bibitem[Vehtari et~al.(2015)Vehtari, Gelman, and Gabry]{vehtari2015pareto}
A.~Vehtari, A.~Gelman, and J.~Gabry.
\newblock Pareto smoothed importance sampling.
\newblock \emph{arXiv preprint arXiv:1507.02646}, 2015.

\bibitem[Wang(2014)]{wang2014importance}
S.~Wang.
\newblock Importance sampling including the bootstrap.
\newblock \emph{Wiley StatsRef: Statistics Reference Online}, 2014.

\end{thebibliography}

\end{document}